\documentclass[global,twocolumn]{svjour}
\usepackage{graphicx}
\usepackage{latexsym}
\journalname{myjournal}

\begin{document}

e+source_ERL_appb-v4a.dvi keepcomments

\title{High-Intensity and High-Brightness Source of Moderated Positrons Using 
       a Brilliant $\gamma$ Beam}
\author{C. Hugenschmidt\inst{1}, K. Schreckenbach\inst{1},
        D. Habs\inst{2,3} \and P.G. Thirolf\inst{2}}
\offprints{Christoph.Hugenschmidt@frm2.tum.de}      
\institute{$^1$ FRM II and Physik Department E21, Technische Universit\"at 
                M\"unchen, 85747 Garching, Germany \\
           $^2$ Fakult\"at f\"ur Physik, Ludwig-Maximilians Universit\"at, 
                85748 Garching, Germany \\
           $^3$ Max-Planck-Institut f\"ur Quantenoptik, 85748 Garching, Germany}
\date{Received: date / Revised version: date}

\titlerunning{High-Intensity and High-Brightness Source of Moderated 
              Positrons Using a Brilliant $\gamma$ Beam}
\maketitle

\begin{abstract}

Presently large efforts are conducted towards the development of highly brilliant
$\gamma$ beams via Compton back scattering of photons from a high-brilliance
electron beam, either on the basis of a normal-conducting electron linac
or a (superconducting) Energy Recovery Linac (ERL).
Particularly ERL's provide an extremely brilliant electron beam, thus enabling
the generation of highest-quality $\gamma$ beams.
A 2.5\,MeV $\gamma$ beam with an envisaged intensity of $10^{15}$photons\,$s^{-1}$, as ultimately
envisaged for an ERL-based $\gamma$-beam facility, narrow band width ($10^{-3}$), and extremely low emittance ($10^{-4}$mm$^2$mrad$^2$) 
offers the possibility to produce a high-intensity bright polarized positron beam.
Pair production in a face-on irradiated W converter foil (200\,$\mu$m thick, 
10\,mm long) would 
lead to the emission of $2\cdot 10^{13}$ (fast) positrons per second, which is 
four orders of magnitude higher compared to strong radioactive $^{22}$Na sources 
conventionally used in the laboratory.
Using a stack of converter foils and subsequent positron moderation, a 
high-intensity low-energy beam of moderated positrons can be produced. 
Two different source setups are presented: a high-brightness positron beam  
with a diameter as low as 0.2\,mm, and a  high-intensity beam of $3\cdot10^{11}$
 moderated positrons per second.
Hence, profiting from an improved moderation efficiency, the envisaged positron 
intensity would exceed that of present high-intensity  positron  sources by a 
factor of 100.

\end{abstract}

\section{Introduction}

Currently large efforts are devoted world-wide to the development of highly 
brilliant $\gamma$ beams.
In such a facility, the $\gamma$ beam with low emittance is created by inverse 
Compton scattering of photons, which are provided by a high-power laser, with an 
ultra-relativistic electron beam
either provided by a normal conducting electron linac or an Energy Recovery 
Linac (ERL).
Until about 2018, it is envisaged to generate a $\gamma$ beam with an intensity of 
$10^{15}$ $\gamma$-photons per second 
(the term \textit{intensity} is used throughout this paper in units of ' number of particles or photons per second') 
and the energy of 2.5(5)\,MeV~\cite{Haj08,Hay10}.
Using a brilliant $\gamma$ beam, positron-electron pairs can be produced in a 
suitable target by pair production. 
A well designed positron source would hence allow to create a moderated positron 
beam of high intensity and/or high brightness.
In addition, the brightness can be further enhanced by positron remoderation.
 
Positron beams are usually generated by using $\beta^+$ emitters such as $^{22}$Na 
and a thin W foil or solid Ne as moderator with an intensity of about 
$5\cdot10^4 -5\cdot10^6$ moderated positrons per second.
At large-scale facilities, such as electron linacs or nuclear reactors, positron 
beams are created with higher intensity by pair production.
At present, the NEutron induced POsitron source MUniCh (NEPOMUC) provides the 
world highest intensity of $9\cdot10^8$ moderated positrons per 
second~\cite{Hug08b}.

In general, various $\gamma$ sources used for pair production such as 
bremsstrahlung targets at linacs, fission $\gamma$'s at reactors or the 
de-excitation of nuclear states emit $\gamma$ radiation isotropically.
For this reason, at present linac or reactor-based positron sources, the large 
area of the converter and positron moderators is the main drawback for improving 
the brightness of the  positron beam. 
Consequently, one can greatly benefit from a low-emittance $\gamma$ beam, 
which allows the adaptation of a converter and positron moderator in an efficient 
positron source geometry.
A brilliant 2.5\,MeV $\gamma$ beam with an envisaged intensity of $10^{15}$photons\,$s^{-1}$ 
would allow to create a positron beam whose intensity exceeds that of 
present high-intensity positron sources by more than two orders of magnitude.

In this paper, various positron source designs and the relevant features are 
discussed.
In particular, two layouts, which provide a high-brightness or a high-intensity 
positron beam, are presented and quantitatively compared with the NEPOMUC beam.

\section{High-brilliant $\gamma$ sources}

High-quality energetic photon beams are versatile tools for a wide range
of physics studies, ranging from precisely probing nuclear properties 
and processes to serving as a starting point for secondary sources such 
as neutrons or positrons.
In general, $\gamma$ beams are produced via Compton back-scattering of
laser photons from a relativistic electron beam.
The presently world-leading facility for photonuclear physics is the 
High-Intensity $\gamma$-ray Source (HI$\gamma$S) at Duke University (USA).
It uses the Compton back-scattering of photons, provided by a high-intensity 
Free-Electron Laser (FEL), in order to produce a brilliant $\gamma$ beam.
The $\gamma$ intensity in the energy range between 1 and 3\,MeV amounts 
to $10^{8}$photons\,$s^{-1}$ with a band width of about 5$\%$~\cite{Wel09}.
Based on a normal-conducting electron linac, the brilliant Mono-Energetic 
Gamma-ray (MEGa-ray) facility at Lawrence Livermore National Laboratory (USA) will 
yield already in 2012 a $\gamma$-intensity of $10^{13}$photons\ ,$s^{-1}$ with an
energy band width of $\leq 10^{-3}$~\cite{Bar10megaray}.
Using similar accelerator technology, at the upcoming Extreme Light Infrastructure -
Nuclear Physics (ELI-NP) facility in Bucharest, until 2015 a $\gamma$ beam will 
become available, providing about the same $\gamma$-intensity and band width in the 
energy range of 1-19\,MeV~\cite{ELI-NP10}. \\
At present, great efforts are also invested all over the world to realize highly 
brilliant $\gamma$ beams based on the Energy Recovery Linac (ERL) technology.
The Energy Recovery Linac (ERL) is a new type of superconducting electron accelerator 
that provides a high-brilliant, high-intensity electron beam. 
The main components of an ERL are an electron injector, a superconducting linac, 
and an energy recovery loop. 
After injection from a high-brilliant electron source, the electrons are 
accelerated by the time-varying radio-frequency field of the superconducting linac.
The electron bunches are transported through a recirculation loop and are re-injected
into the linac during the decelerating RF phase of the superconducting cavities.
So the beam dump has to take only a small fraction of the beam energy. 
In this way, the energy is recycled very efficiently and re-used to accelerate 
a new bunch of electrons.
ERL's create high-energy, high-brilliant $\gamma$ beams by Compton 
back-scattering of photons from high-energy (0.1-5\,GeV) electrons, 
recirculating the photons in a very high-finesse cavity with MW power.
ERL technology has been pioneered at Cornell University (together with the 
Thomas Jefferson National Laboratory)~\cite{Bil10a,Bil10b,liepe10},
where an ERL is presently constructed for a 5\,GeV, 100 mA electron beam.
At the KEK accelerator facility in Japan, an ERL project is presently pursued
aiming at a $\gamma$ beam with an intensity of $10^{13}$photons\,$s^{-1}$~\cite{Haj08,Hay10}. 
In Germany, a high-current and low-emittance demonstrator ERL facility (BERLinPro) 
is developed at the Helmholtz Zentrum Berlin~\cite{Jan10}.
Three different operation modes are conducted: high-flux mode, 
high-coherence (brilliance) mode, and a short-pulse mode~\cite{Bil10a,Bil10b,Hay10}.
For our purpose of positron production, the high-flux mode is of particular interest.
Moreover, the facility can be optimized to the specific needs of the intended
application. When, e.g., as in the present case aiming at the production of a high-brightness positron beam, 
a small $\gamma$ beam spot size and low beam divergence is 
more important than the superb energy band width provided by an ERL. 
The ultimately envisaged photon intensity is $>10^{15}$photons\,$s^{-1}$ in an energy range of 
0.5-20\,MeV. 
Such a facility would provide a brilliant pulsed (ps pulse length) $\gamma$ 
beam with a narrow band width of about $< 10^{-3}$ and a low emittance 
of $10^{-4}mm^2mrad^2$.  

\section{Positron production by a high-brilliant $\gamma$ 
           beam}
\subsection{Principle of the positron source}
There are two fundamentally different setups for the creation of a moderated 
positron beam using a brilliant $\gamma$ beam.
Either the $\gamma$-positron-electron converter and the positron moderator are 
separate components, or the converter is used as positron moderator as well, and 
hence the moderated positrons are extracted directly from the converter surface.
The production and subsequent moderation in the same component is 
called \textit{self-moderation}.
In order to create a bright positron beam, a moderator  should be used with 
high efficiency and narrow band width of the emitted positrons.
However, the choice of the applied moderator material strongly depends on the 
final source layout. 
Various designs specifically suited for brilliant $\gamma$ beams and the 
respective features are presented in Section\,\ref{geometry}.

\begin{figure}
 \includegraphics[width=1.\columnwidth]{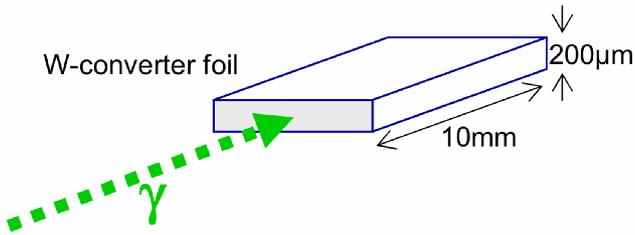}
   \caption{Sketch of a simple W converter foil irradiated with a brilliant $\gamma$ beam for positron-electron pair production.  
   \label{plate}}
 \end{figure}

The $\gamma$ conversion into positron-electron pairs takes place in a material 
with high nuclear charge $Z$, such as Pt or W (also suitable moderator materials), 
since the pair production cross 
section $\sigma_{PP}$ increases approximately proportional to $Z^2$.
At a $\gamma$ energy of 2.5\,MeV, the pair production cross section $\sigma_{PP}$ 
for Pt and W  amounts to 2.386 and 2.713 barn/atom, respectively.
In addition, the converter material should have a high melting temperature due 
to the high local heat dissipation. 
The optimum thickness of a single converter foil with highest amount of emitted 
positrons is in the range of 0.4-0.5\,g/cm$^2$~\cite{Kru90}.
In order to create free (fast) positrons, one could simply choose a thin W 
converter foil (density 19.35 g/cm$^3$, e.g., 200\,$\mu$m thick, 
10\,mm long), which is irradiated by 
the $\gamma$ beam on the face side -as sketched in Figure\,\ref{plate}- leading to a  $\gamma$ absorption of about 
55.4\,$\%$.
The amount of free fast positrons can be calculated, considering the pair 
production cross section $\sigma_{PP}$ and the probability for fast positron 
emission from a 200\,$\mu$m thick W foil, which amounts to 20$\%$~\cite{Kru90}.
Thus, a $\gamma$ beam with a $\gamma$ intensity of $10^{15}$photons\,$s^{-1}$ would lead to 
the emission of $2\cdot 10^{13}$ (fast) positrons per second from an area of about 
2\,mm$^2$.
The positron intensity of this source would be four orders of magnitude higher 
than that from strong radioactive $^{22}$Na sources (2\,GBq) conventionally used in 
the laboratory.

As shown below, the fraction of moderated and emitted positrons can be enhanced by 
using a stack of several converter foils. 
Suited materials for positron moderation are metals with negative positron work
 function $\Phi^+$ such as Pt ($\Phi^+=-1.95$\,eV \cite{Hug02d}) and 
W ($\Phi^+=-3.0$\,eV \cite{Coleman2000}) or solid rare gases \cite{Mil86}.
The moderation efficiency of W is known to be higher than that of Pt and amounts 
to about $4\cdot10^{-4}$~\cite{Lyn85}.
However, depending on the surroundings, Pt might become more reliable during 
operation due to the in-situ annealing of radiation-induced defects. \cite{Hug04a}
Solid rare gas moderators exhibit higher moderation efficiencies, but the 
bandwidth of the resulting positron beam is larger due to the emission of 
epithermal, i.e., not fully thermalized positrons.
The  moderation efficiency  was measured with a 50\,$\mu$m solid Ne film on top 
of a $^{22}$Na source and amounts to $\epsilon_{mod}=3\cdot10^{-3}$ \cite{Mil86}. 
The energy spread of the Ne moderated positrons was found to be 0.58\,eV, and 
hence, about one order of magnitude worse than that from a W moderator \cite{Mil86}.
In general, the comparison of the moderation efficiencies is often difficult, since 
it does not only  depend on the primary positron spectrum, but, even more 
importantly, on the used moderator geometry. 
For this reason, efforts were done to increase the yield of moderated positrons 
by optimizing the source-moderator layout. 
In the following, the moderation efficiency is defined as the number of moderated 
positrons in the slow positron beam divided by the number of produced (fast) 
positrons in the converter. 
In addition, remoderation of the positron beam would lead to a further enhancement 
of the beam brightness (see e.g. \cite{Mil80c,Pio08}). 

\subsection{Geometry of the converter-moderator setup}
\label{geometry}
\begin{figure*}
 \includegraphics[width=1.8\columnwidth]{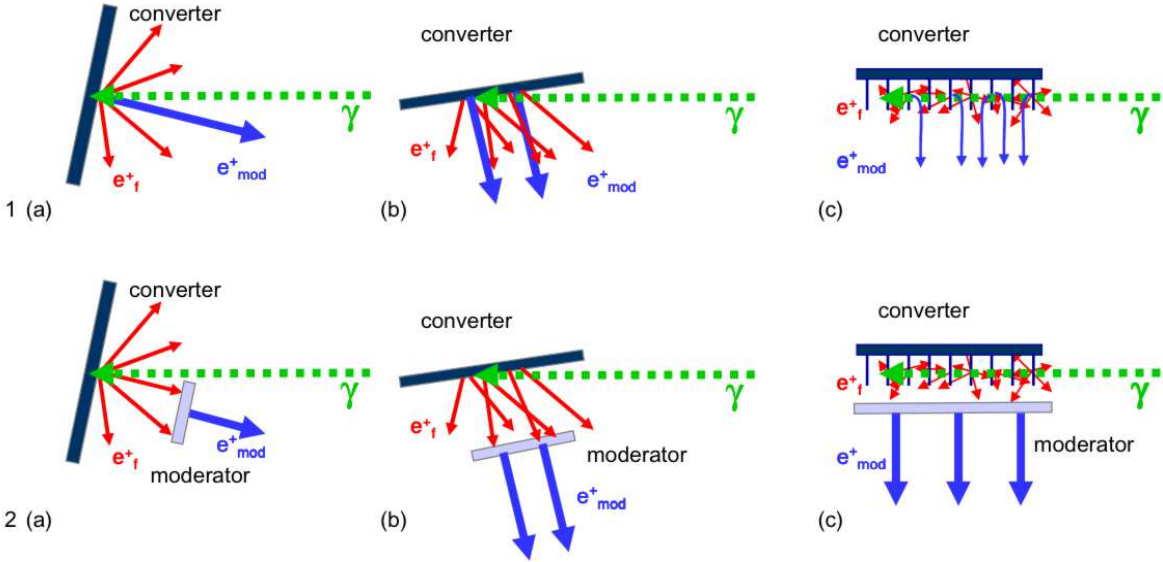}
   \caption{Schematical view of various converter-moderator layouts using a 
            brilliant $\gamma$ beam. (1) Conversion and moderation in the same 
            component 
       (self-moderation), and (2) moderator separated from the converter. The 
       features of the versions (a)-(c) are discussed in the text. 
   \label{design}}
 \end{figure*}

An overview of two basic layouts specifically suited for a brilliant $\gamma$ 
beam-induced positron source, with three different configurations each, is given
 in Figure\,\ref{design}.
In the first layout (1) the $\gamma$-positron-electron conversion and the 
positron moderation take place in the same component (self-moderation).
The second one (2) consists of the converter and a separated positron moderator.
  
In the layout (1), the application of a metallic converter and moderator seems 
to be most convenient due to the high local heat load.
Therefore, a converter material, such as W or Pt, should be applied in order to 
operate the converter reliably.
In general, using the geometry (2), the moderator has to be mounted as close as 
possible to the converter in order to increase the solid angle for positron 
moderation. 
In this layout, solid rare gases can be applied, e.g., a thin layer of Ne frozen
 on top of a Be foil, which would lead to a higher moderation efficiency, but 
also to a higher band width of the moderated positron beam and a larger beam 
diameter.

(a) The tiniest beam spot, and probably the highest brightness, is achieved 
using layout 1(a), since the diameter of the moderated positron beam is 
barely larger than the interaction area defined by the $\gamma$ beam. 
The usage of a separate Ne moderator (setup 2(a)) would increase the moderation 
efficiency  at the expense of a lower solid angle for the positron irradiation. 
 
(b) The grazing incident $\gamma$ beam shown in 1(b) and 2(b) would increase 
the total rate of produced positrons, but the positron beam area would be that of a 
largely elongated ellipse with lower brightness compared to versions 1(a) and 2(a). 
 
(c) In the layouts 1(c) and 2(c), the converter consists of a stack of thin 
W or Pt foils, which would lead to a very efficient absorption of the 
$\gamma$ beam.
The positron production and emission rate can be improved with the number of 
foils, i.e., the cumulated thickness of the converter material.  
Due to the narrow $\gamma$ beam, short foils could be used facilitating the 
extraction of the moderated slow positrons.
In version 1(c), a suitable electrical acceleration field has to be applied in 
order to extract the moderated positrons since they are emitted perpendicular 
to the moderator surface.
This challenge will be overcome in version 2(c), where a moderator can be placed 
close to the converter.
The beam extraction could also be performed perpendicular to the plane of 
projection for the setups 1(c) and 2(c). 
However, similar to the layouts 1(b) and 2(b), the cross section of the resulting 
positron beam would be largely elongated in one dimension. 

\subsection{The high-brightness and the high-intensity positron beam}

In summary, we propose to focus on two most promising source setups, which 
should be realized in a brilliant $\gamma$ beam facility: 
The first one would generate a \textit{high-brightness} (HB), and the second 
one a \textit{high-intensity} (HI) moderated positron beam.

The HB source geometry corresponds very much to the layout 1(a) in 
Figure\,\ref{design}.
In the thin layer limit, i.e., low  $\gamma$ absorption, the production rate of 
(fast) positrons $R^+$ can be approximated by $R^+=I_{\gamma} \cdot \sigma_{PP} 
\cdot n_W \cdot d_W$ with the $\gamma$ intensity $I_{\gamma}$ and the W atom 
density $n_W$.
For a W foil with a thickness of $d_W$=250\,$\mu$m, the fraction of emitted 
positrons amounts to $f^+_{em}\approx 0.2$ with respect to the number of 
produced positrons and $f^+_{em}$ is much higher for thinner foils, e.g., $f^+_{em}\approx 0.93$  for 
$d_W$=10\,$\mu$m \cite{Kru90}. 
Hence, using a 250\,$\mu$m W foil in back reflexion geometry for the HB source, 
about 20$\%$ of the produced positrons can contribute to the emission of 
moderated positrons.
The fraction of fast positrons $f^+_{st}$ with a mean energy of 750\,keV stopped 
in a surface layer of 50\,nm amounts to  $f^+_{st}\approx1.8\cdot10^{-4}$.
According to the positron diffusion length in W of 135\,nm \cite{Veh83}, it is 
assumed that almost all positrons thermalized in the 50\,nm surface layer 
reach the surface.
Accounting for losses at the surface due to Positronium formation $f_{Ps}$, and
 trapping in surface states $f_{surf}$, the positron probability to be emitted 
as a moderated positron is $p_{mod}=1-(f_{Ps}+f_{surf})\approx 0.4$.
This consideration and the value for $p_{mod}$ is in agreement with the moderation 
efficiency experimentally determined for W(100)  using a $^{22}$Na source with an 
according mean positron energy of 200\,keV \cite{Lyn85}.
Thus, the yield of moderated positrons $Y^+_{mod}$ is calculated as
 $Y^+_{mod}=R^+ \cdot f^+_{em} \cdot f^+_{st} \cdot p_{mod}$.
With the numbers given above, one obtains $Y^+_{mod} \approx R^+ \cdot 
1.5\cdot 10^{-5}$.

The positron beam diameter is slightly greater than the $\gamma$ beam diameter.
Its increase is of the order of the mean positron range of about 0.1\,mm. 
The positron diffusion length is about three orders of magnitude lower and 
hence negligible.
The parameters expected for a HB positron beam are shown in 
Table\,1. 
Besides the higher brightness, a major advantage of the HB source is the 
relatively simple setup, where an electrical extraction field has to be 
applied for positron acceleration.

\begin{figure*}
 \includegraphics[width=1.8\columnwidth]{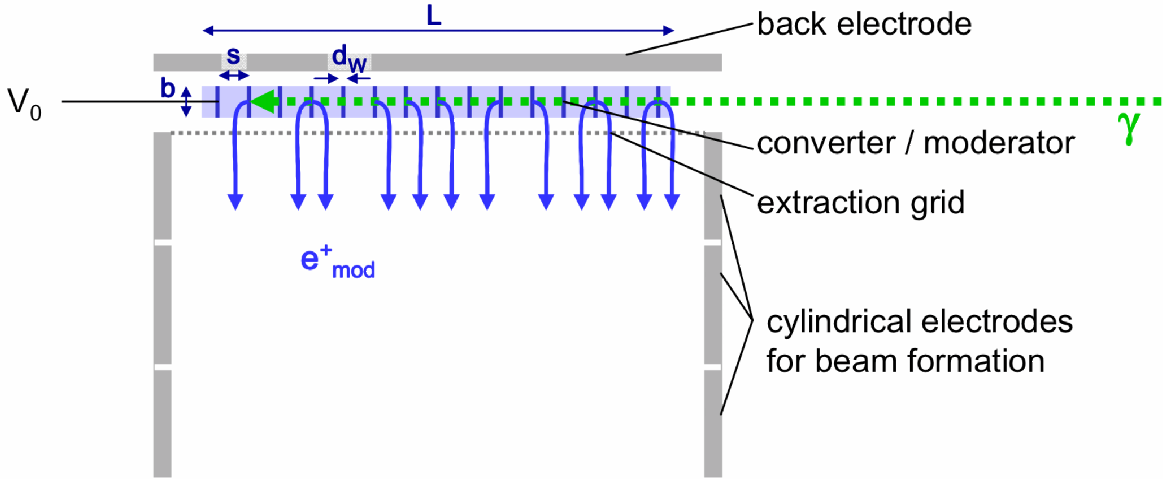}
 \caption{Scheme of a converter-moderator configuration irradiated by a brilliant 
     $\gamma$ beam  for the generation of a high-intensity moderated positron beam. 
     The converter-moderator itself consists of a stack of $N$ single crystalline 
     W foils of thickness $d_{W}$. The ratio of the foil width $b$, which is 
     in the order of the diameter of the $\gamma$ beam, and the spacing $s$ between 
     the foils is 3:1.
     The total length $L$ is hence given by $L\approx N(s+d_{W})$.  The total setup 
     consists of the converter-moderator block (on high potential V$_0$) which is mounted between a back electrode on 
     higher potential and an acceleration grid in order to extract the moderated 
     positrons. (Cylindrical) electrodes are used for beam formation.
\label{design_best}}
 \end{figure*}

In the following, we present a more detailed source geometry for the creation 
of a HI positron beam.
The layout of the HI source shown in Figure\,\ref{design_best}  is similar 
to 1(c) shown in Figure\,\ref{design}, and can easily be extended to the 
version 2(c).
The converter-moderator, which is operated in the vacuum, consists of a stack of $N$ single crystalline W foils of 
thickness $d_{W}$ with a spacing between the foils of $s$.
The width $b$ of the W foils would be of the order of the diameter of 
the $\gamma$ beam.
The length of the foils (perpendicular to the drawing plane of 
Figure\,\ref{design_best}) can be chosen much larger than $b$ to facilitate 
the extraction of the moderated positrons. 
In order to keep the total length of the converter not too long, a ratio 
of $b$:$s$=3:1 is expected to be reasonable for a good enough beam extraction by 
an electric field.
Such a converter can be either set up by using $N$ W(100) foils clamped between 
small W holders, or the whole component is cut out from a long W single crystal 
using a laser cutter or spark erosion. 
Tilting of the equally spaced foils would increase the effective absorption 
length in the foils, but --keeping the ratio $b$:$s$ constant-- the spacing by 
the same amount as well.
Hence, at a given total absorption length, the number of foils $N$ would 
decrease, and the length of the whole converter would not change.

In the following, the arrangement with parallel foils, i.e., 
perpendicular to the $\gamma$ beam, is presented.
The advantages of this setup are higher mechanical stability, lower heat input 
per foil, better usage of reflected fast positrons, and higher solid angle for 
the individual foils with respect to the neighboring ones, leading to a higher 
efficiency to produce moderated positrons.   

The converter-moderator block is aligned in direction of the $\gamma$ beam which 
interacts with the W foils by pair production.
In contrast to the primary produced fast positrons, the moderated  positrons
 are emitted perpendicular to the W(100) surface.
Since their  primary kinetic energy amounts to 
$E^+_{mod}=-\Phi^+=2.8\,eV$\cite{Jac90}, an electrical extraction field is needed, 
which is provided by the back electrode and the extraction grid as shown in 
Figure\,\ref{design_best}.
The potential $V_0$ applied at the converter-moderator block defines the  final 
kinetic energy of the positron beam $E^+_{kin}=eV_0-\Phi^+$. 
The beam should be extracted in a zero magnetic field in order to maintain the 
low primary divergence and the high grade of polarization of the moderated 
positron beam.

In order to estimate the resulting moderated positron yield $Y^+_{mod}$ from the 
production rate $R^+$, we first consider a single foil, and a two-foil arrangement
 in the extreme limits:
A single converter foil with two surfaces emitting moderated positrons would give:
$Y^+_{mod}=R^{+}\cdot 2\cdot f^+_{em}\cdot f^+_{st}\cdot p_{mod}$.
For a stack of thin foils with $s<<b$ and for not too high total converter 
length, i.e., not much longer than the mean positron range, a produced positron 
could be stopped in any foil and has a certain probability to reach the surface.  
In this case, $f^+_{st}$ can be approximated by the constant value given above, 
and $Y^+_{mod}$ would just scale with the number of positron emitting 
surfaces $2N$:
$Y^+_{mod}=R^{+}\cdot 2N \cdot f^+_{em}\cdot f^+_{st}\cdot p_{mod}$.
In the second limit for $s>>b$, each foil would act independently, and hence, 
$Y^+_{mod}$ would be the sum of the $N$ producing and moderating foils with 2 
surfaces each:
$Y^+_{mod}= 2N \cdot (R^{+}/N) \cdot f^+_{em}\cdot f^+_{st}\cdot p_{mod}$.

In a realistic arrangement, the calculation of $Y^+_{mod}$ becomes very 
complicated, since each W production foil irradiates the others by positrons with 
a certain probability.
Hence, the respective solid angles, positron attenuation, and stopping in  a 
surface layer has to be considered. 
Note, that $f^+_{st}$ has to be calculated as function of the positron energy, 
which depends on the respective positron absorption length  $f^+_{st}(E(x))$, 
since it might become considerably higher for low-energy positrons. 

We propose a converter of $N=2500$ parallel W foils with $b$=200\,$\mu$m, a
ratio of $b$:$s$=3:1 and hence spacing s=67\,$\mu$m, and $d_{W}=10\,\mu m$.
Thus, the converter length would amount to $ L\approx 192\,mm$, and the cumulated 
thickness of the W converter material would be $N\cdot d_{W}=25$mm.
The total $\gamma$ absorption in the W foils would be about 87\,$\%$, and 
consequently, the corresponding heat input due to $\gamma$ heating of 350\,W 
has to be dissipated by a cooling device at the converter.
The total positron production rate would be $R^+=1.62\cdot 10^{14} s^{-1}$. 
Due to the $\gamma$ absorption, the number of produced positrons in an individual W foil would decrease with the foil number N .   
The yield of moderated positrons of the whole HI setup can be estimated 
by $Y^+_{mod}=2N\cdot (R^{+}/N) \cdot f^+_{em}\cdot f^+_{st}\cdot p_{mod}\cdot 
(1+\eta)$.
The term $\eta$ accounts for an additional contribution of the $N-1$ other 
foils to the emission of moderated positrons of each single foil.
Hence, $\eta$ can reach values well above 1, since  $f^+_{st}(E(x))$ becomes
 much larger than $1.8\cdot10^{-4}$ for low-energy positrons.
Taking the numbers given for $f^+_{em}$, $f^+_{st}$, and $p_{mod}$, one 
gets $Y^+_{mod}=2.17\cdot 10^{10} \cdot (1+\eta) s^{-1}$.
Accounting for the respective solid angles, the emitted and slowed-down 
positrons of the neighboring foils lead to the additional emission of moderated positrons resulting in $\eta\approx 13$.
Hence, the positron yield can be roughly estimated 
and amounts to $Y^+_{mod}=3\cdot 10^{11} s^{-1}$.
Note, that this value can even be higher due to the contribution 
of reflected positrons, which are moderated, and a higher moderation efficiency 
of inelastically scattered positrons.
Thus, compared to the HB setup, the slow-positron yield would be about 3500 
times higher at the HI source, and it would exceed the intensity of the 
upgraded NEPOMUC \cite{Hug11x} source by two orders of magnitude.
Due to the much larger beam spot, which is expected to be about 400\,mm$^2$, 
the brightness of the HI beam would be lower than that for the remoderated beam at 
NEPOMUC, and more than two orders of magnitude worse than that at the HB source 
(see Table\,1).
Note, that, taking into account the longitudinal energy spread, the brightness
 of the HI or the HB pulsed beams (pulse length of a few ps) could be enhanced 
considerably by narrowing the energy width at the expense of time resolution. 

Besides the considerations with respect to the positron production rate and yield 
of moderated positrons, other factors have to be considered as well, such as 
converter cooling, 
annealing of the moderator, and positron beam extraction.
Independent of the source layout, the moderator --or the converter if it acts 
as moderator as well-- has to be floated on a variable potential in the range 
of 0.01-5\,kV (or even higher) in order to adapt the kinetic energy of the 
positrons to the experimental requirements.
Additional lenses have to be mounted in front of the moderator for positron 
beam formation, and the positron beam should be magnetically guided to the 
experimental setups.

In order to estimate various positron beam parameters such as 
$R^+$, $Y^+_{mod} $, diameter $d^+$, and brightness $B$, we assume the 
availability of a brilliant $\gamma$ beam with an intensity 
$I_{\gamma}$=10$^{15}$photons\,$s^{-1}$, an energy of 2.5(5)\,MeV, and a diameter 
of 0.1\,mm.
According to Liouville's theorem, the product of the divergence, the beam 
diameter, and the longitudinal component of the momentum $\sqrt{2mE_L}$ is 
constant.
Hence, the brightness $B$ defined as $B=\frac{I}{\theta^2 d^{+2} E_L}$ is a 
good figure of merit for a positron beam of intensity (particles per second) $I$, diameter $d^+$, 
divergence $\theta=\sqrt{E_T/E_L}$ with transversal and longitudinal components 
of the positron energy $E_T$ and $E_L$. 
Note, that this definition of $B$ is commonly used for the characterization of
 positron beams (see e.g. \cite{Coleman2000,Mil80c,Sch88}).
However, in the literature the terms\textit{ brilliance} and \textit{brightness} 
are not used in a consistent way. 

Here, we assume that all moderated positrons leaving the foils are extracted, 
i.e., $I=Y^+_{mod}$, 
and the kinetic energy of the positrons is set to $E_L$=1\,keV.
All parameters are calculated for both the HB and the HI positron source as well,  
and summarized  in Table\,1. 
Since the HI source would provide a beam cross section, which is largely  
elongated in one dimension, an effective diameter of a circular shaped beam
 spot of the same size is given.
For comparison, the values for the NEPOMUC beam and its upgrade  are shown as 
well.
As a result, using the HB source, one can expect a positron beam with a
 brightness, which is more than two orders of magnitude higher than that of 
the remoderated NEPOMUC beam.
With the HI layout, the positron intensity is expected to be two orders of 
magnitude higher than that of the NEPOMUC upgrade.
Depending on the properties of the $\gamma$ beam, these parameters scale with 
the available $\gamma$ intensity.
An additional remoderation device could be used for further brightness enhancement.

\begin{table*}[ht]
\begin{tabular}{l|c|c|c}
    {\bf } &     {\bf} & \multicolumn{ 2}{c} {\bf brilliant $\gamma$ beam~~~~~~~~}  \\

    {\bf } &     {\bf NEPOMUC} & \ {\bf HB source} & \ {\bf HI source} \\
\hline
 setup & Pt 125\,$\mu$m trans.  & W 250\,$\mu$m refl. & W 2500 x 10\,$\mu$m refl. + trans.\\
 $R^+$ $[s^{-1}]$ &  2.6.$\cdot 10^{14}$ (5.9.$\cdot 10^{14})$ & 5.7$\cdot 10^{12}$ 
                                                              & 1.6$\cdot 10^{14}$ \\
 $Y^+_{mod}$ $[s^{-1}]$ & 9.0$\cdot 10^{8}$ (3.0$\cdot10^{9}$) & 8.5$\cdot 10^{7}$ 
                                                               & 3.0 $\cdot 10^{11}$ \\
 $\epsilon_{mod} $  & 3.5$\cdot 10^{-6}$ (5.1$\cdot10^{6}$) & 1.5$\cdot 10^{-5}$ 
                                                           &  1.9 $\cdot 10^{-3}$  \\
 $E_T$ $[eV]$ 	&   50$^*$ (0.15) & 2.5$\cdot 10^{-2}$ & 3.0  \\
 $\theta$ $[mrad]$ & 2.2$\cdot 10^{-1*}$ (1.2$\cdot 10^{-1}$) & 5.0$\cdot 10^{-3}$ 
                                                             & 5.5$\cdot 10^{-2}$ \\
 $d^+$ $[mm]$ &   7$^*$  (70) &   0.2 &   23 \\
 $B$ $[(mrad^2 mm^2 eV s)^{-1}]$  & 3.7$\cdot 10^{5*}$, 5$\cdot 10^{8}$ remod. 
                                  & 8.5$\cdot10^{10}$ & 1.9$\cdot 10^{8}$ \\ 
	 &  (4.1$\cdot10^{6}$)  &  &    \\
 spin polarization &   no 	&   yes &  yes \\
 operation mode	   &   continuous &   pulsed ($\sim$ps)  &  pulsed ($\sim$ps)  \\

\label{Table:parameter}
\end{tabular}  
\caption{Expected beam parameters for a HB and HI positron source using a 
high-brilliant $\gamma$ beam with $I_{\gamma}$=$10^{15} s^{-1}$ in comparison 
with the NEPOMUC source.
The numbers given in parentheses refer to the NEPOMUC upgrade in 2011 
\cite{Hug11x}.
$R^+$ positron production rate, $Y^+_{mod}$ yield of moderated positrons, 
i.e. positron beam intensity, $\epsilon_{mod}$ moderation efficiency, $E_T$ 
transversal energy, $\theta$ divergence, $d^+$ positron beam diameter, and $B$ brightness 
of the positron beam at $E_L$=1\,keV.
($^*$measured values at the first accessible position of NEPOMUC.)}
\end{table*}

In general, the key features of a low-energy positron beam based either on the HB or HI 
layout using a high-brilliant $\gamma$ beam would be the following:

\begin{itemize}
\item \textbf{$\gamma$ energy:} 
The energy of the  $\gamma$ beam can be varied  in the range of a several MeV in 
order to maximize the positron production and emission rate as well as the yield 
of moderated positrons. 
\item \textbf{Band width:} 
Due to the small band width of the $\gamma$ beam, no unwanted $\gamma$'s are 
produced with $E<2mc^2$ which do not contribute to the pair production. 
Therefore, the heat load compared to linac or reactor based positron sources 
is expected to be considerably lower.
\item \textbf{Diameter and brightness:} 
The intrinsic small diameter of the $\gamma$ beam leads to an accordingly small 
positron beam. 
Dependent on the source geometry, a higher brightness of the moderated beam is 
expected as well.
\item \textbf{Polarization:}
Using a switchable fully polarized $\gamma$ beam, a spin-polarized positron
 beam can be created. 
Since the positron polarization is almost entirely maintained during 
moderation~\cite{Zit79}, spin-dependent experiments may become feasible. 
\item \textbf{Time structure:} 
The time structure provided by the pulsed $\gamma$ beam is barely deteriorated 
by the moderation process since positrons thermalize very rapidly (within a few 
picoseconds) after production or implantation.
It is expected that a smearing of the beam pulse is mainly caused by the 
resulting positron spectrum, different flight paths in the source and position 
dependent acceleration of the moderated positrons.
However, the usefulness of  the time structure of the $\gamma$ beam strongly 
depends on the positron beam application, e.g., for coincidence techniques 
using lasers rather than for positron lifetime spectroscopy.
\item \textbf{ Access:} 
The source area of the $\gamma$ beam will be easily accessible. This would 
facilitate the change of the source setup considerably. 
For future applications, we recommend to install a \textit{source switch} in 
order to allow a quick change from a high-brightness to a high-intensity 
positron beam.
\item \textbf{Radiation field} 
Due to the well defined relatively low energy of the $\gamma$ beam, e.g., 
2.5(5)\,MeV, the creation of radiation induced defects is expected to be lower 
than that at positron source setups using brems\-strahl\-ung targets at linacs or 
$\gamma$ rays produced at nuclear reactors. In addition, no radioactivity is 
created by activation. 
\end{itemize}

\section{Outlook and conclusion}

\subsection{First $\gamma$ beam based positron sources}

Great efforts are presently made to develop high-brilliant $\gamma$ beams.
Within the next years, two $\gamma$-beam facilities with an intensity of 
$10^{13}$photons\,$s^{-1}$ will become available.  
Both, the Mono-Energetic Gamma-ray (MEGa-ray) source in Livermore
(commissioning starting in 2012), and the 
Extreme Light Infrastructure - Nuclear Physics facility (ELI-NP) planned in 
Bucharest (operation envisaged for 2015), designed for $10^{13}$ photons\,$s^{-1}$,
would be suited to install a $\gamma$-beam based positron source,
potentially exceeding the presently strongest positron source (NEPOMUC) by 
about a factor of three.
Feasibility studies for the positron beam production, using the source layouts 
as presented here, can be performed in advance, and experimental data can be 
gained already at much lower $\gamma$ beam intensity.
It is expected that even more brilliant (ERL-based) $\gamma$ beam facilities 
will become operational within the next decade with an ultimate intensity 
of $10^{15}$photons\,$s^{-1}$ and a diameter of 0.1\,mm. 

We propose the installation of both the HB and the HI positron source in the 
target area at ELI-NP.
The low-energy positron beam can be transported over long distances and through 
bends without intensity loss or considerable deterioration of the positron beam 
quality as long as the positrons are guided adiabatically in a static 
homogeneous magnetic field.
There are two main techniques to realize the homogeneous longitudinal magnetic 
guide field: 
Either solenoid coils directly mounted on the beam line or a Helmholtz-like 
setup of several coils with larger diameter.
Additional saddle coils are  required in order to compensate for transversal 
field components, and $\mu$-metal shielding can be mounted as well.
Therefore, the moderated positrons created at ELI-NP can be guided to an external
 experimental area if the place close to the target is limited.   

After calculation of several entities such as production and emission rates of 
positrons for various converter materials and different geometries and
simulation of positron beam trajectories, experimental data have to be gained 
in order to optimize the positron source setup.
Such experiments can also be performed at a low-flux  $\gamma$ beam facility.
Afterwards, the optimized HB and HI positron sources can be installed where  
brilliant $\gamma$ beams become available at ERL's.
 
In the following several aspects are considered for first experiments:

\begin{itemize}
\item The energy dependent pair production cross section increases considerably
 with increasing $\gamma$-energy leading to a higher positron production rate. 
However, the slow-positron yield does not increase in an analogous manner, since 
the positron moderation efficiency decreases with higher energy.
Therefore, the positron yield as function of $\gamma$ energy should be determined.
 \item Several converter geometries can be compared in order to increase the 
intensity and/or the brightness of the moderated positron beam.
A higher mass of the converter, i.e., thicker foils and/or more foils, would lead 
to a higher positron production rate.
An increased surface-to-volume ratio would result in a higher yield of moderated 
positrons.
Using a separate moderator, the solid angle with respect to the converter should 
be maximized in order to extract as many positrons as possible.
\item Two setups should be compared and optimized for positron beam applications: 
the HB setup and the HI layout based on self-moderation or  with a separated Ne 
moderator.
For various setups the spectrum and the brightness of the slow-positron beam have 
to be determined experimentally.
\end{itemize}

\subsection{Future applications of a high-intensity positron beam}

Depending on the experimental requirements, a bright beam with a diameter of 
about 200\,$\mu$m as delivered from the HB source might be more suited than the 
high-intensity beam from the HI source.
However, one could use an additional remoderation device to  further enhance the 
brightness.

There are many applications, which would benefit from a strong positron source 
providing a high-intensity low-energy positron beam (see e.g. \cite{Hug10a}).
A high positron intensity would be very advantageous for the generation of 
(re-)moderated positron micro-beams for all scanning beam techniques.
In materials science and solid-state physics, such a micro-beam would greatly 
enhance the statistics for spatially resolved defect spectroscopy, using scanning 
positron lifetime or Doppler-broadening measurements.
For the application of coincidence techniques, a high-intensity positron beam 
is even more important, since the measurement time would be drastically reduced
and the spatial resolution would be improved as well.
Such techniques are Coincident Doppler-Broadening Spectroscopy (CDBS), that 
allows to investigate the chemical environment of open volume defects,
Age-MOmentum Correlation (AMOC), where the positron lifetime and the 
Doppler-shift are detected simultaneously for each annihilation event; or the 
determination of the Angular Correlation of Annihilation Radiation (ACAR) in 
order to study the electronic structure of matter.
A bright intense low-energy beam would allow to further develop Positron 
annihilation induced Auger-Electron Spectroscopy (PAES) for spatially resolved 
surface analysis.
In atomic physics, intense positron beams are desired, since small-diameter beams carrying a high intensity are crucial in all kinds of positron scattering 
experiments.
For the creation of mono-energetic Positronium (Ps) beams and for the Ps$^-$ 
production, a high intensity of the moderated positron beam would be very helpful.
This would hence allow the spectroscopy of Ps and  Ps$^-$. 
In addition, for fundamental experiments, the specific formation of  the Ps$_2$ 
molecule or even the creation of a Ps Bose-Einstein condensate would become 
possible.

\subsection{Conclusion}

With the availability of high-brilliant $\gamma$ sources, the realization of   
high-intensity and high-brightness positron sources will become possible within 
a few years.
The efforts and  costs of such positron sources are expected to be not too 
elaborate.
At a brilliant $\gamma$ beam with an envisaged intensity of 
$I_{\gamma}=10^{15}$photons\,$s^{-1}$, a positron beam would exceed the intensity of the 
upgraded high-intensity positron source NEPOMUC by a factor of 100.
Using the high-brightness setup, the brightness is expected to be more than 
two orders of magnitude higher than that of the present remoderated positron beam at 
NEPOMUC.
In the final configuration,  we recommend the implementation of two  different 
source setups.
Hence, one could choose between a high-brightness positron beam with a tiny 
diameter in the order of 0.2\,mm or a  larger high-intensity beam which 
provides about $3\cdot10^{11}$ moderated positrons per second. 
The availability of such an intense positron source would greatly improve all 
kinds of positron beam applications in material science, solid-state, surface, 
and atomic physics as well as fundamental experiments using positrons or 
positroniums.


\end{document}